\documentstyle[12pt,epsf]{article}
\input epsf.tex

\begin{document}

\thispagestyle{empty}

\vspace{24pt}
\begin{center}

{\large\bf Can Doubly Strange Dibaryon Resonances}\\
{\large\bf be Discovered at RHIC?}\\

\vspace{24pt}

S. D. Paganis,\footnote{Present address: Nevis Laboratories, Columbia
University, P. O. Box 137, Irvington, NY 10533.}
G. W. Hoffmann, R. L. Ray\footnote{Corresponding author: electronic address:
ray@physics.utexas.edu}, J.-L. Tang\footnote{Present address: Department of
Physics, National Central University, Chung-Li, Taiwan 320.} and T. Udagawa   \\

\vspace{12pt}

{\sl Department of Physics}\\
{\sl The University of Texas at Austin, Austin, Texas 78712}\\

\vspace{24pt}

R. S. Longacre \\

\vspace{12pt}

{\sl Physics Department, Brookhaven National Laboratory, Upton, NY  11973}\\

\begin{abstract}
The baryon-baryon continuum invariant mass
spectrum generated from
relativistic nucleus + nucleus collision data  
may reveal 
the existence of
doubly-strange dibaryons not stable against strong decay 
if they lie within a few
MeV of threshold.  Furthermore, since the
dominant component of these states is a
superposition of two color-octet clusters which can be produced
intermediately in a color-deconfined quark-gluon plasma (QGP), an 
enhanced production of dibaryon resonances could be a signal of QGP formation.
A total of eight, doubly-strange dibaryon states are
considered for experimental search using the STAR detector
(Solenoidal Tracker at RHIC)
at the new Relativistic Heavy Ion Collider (RHIC).
These states may decay to
$\Lambda \Lambda$ and/or $p \Xi^-$, depending on the resonance energy.
STAR's large acceptance, precision tracking and vertex reconstruction
capabilities, and large data volume capacity, make
it an ideal instrument to use for such a search.
Detector performance and
analysis 
sensitivity are studied as a function of resonance production rate and width 
for one particular dibaryon which can directly strong decay to $p \Xi^-$,
but not $\Lambda \Lambda$.  
Results indicate that such resonances may be discovered using STAR if
the resonance production rates are
comparable to coalescence model predictions for dibaryon bound states.

\end{abstract}

\end{center}

\vspace{0.75in}
\leftline{To be published in {\it Physical Review C}}

\newpage

\section{Introduction}
\label{section:introduction}

It is generally assumed that quantum chromodynamics (QCD) is capable of
describing the structure and spectroscopy of baryons and mesons.  In addition,
a number of models based on QCD
predict other types of quark and gluonic
systems such as pure gluonic states (glueballs), hybrids (q$\bar{\rm q}$g),
four quark di-mesons (qq$\bar{\rm q} \bar{\rm q}$,
q$\bar{\rm q}$q$\bar{\rm q}$) \cite{pdg}, penta-quark states
(qqqq$\bar{\rm q}$), six quark dibaryons \cite{jaffe,sakai,kling},
strangelets, and for
large systems several fm in size
at sufficiently high temperature and/or net baryon density,
the color-deconfined
quark-gluon plasma (QGP).  Extensive experimental searches have been made
for all these systems.

In particular, many experiments ({\it e.g.}
\cite{sakai,kling,HexpKK,shah,belz}) have
searched for a stable (at least to strong decay), doubly-strange
flavor-singlet dibaryon (H$_0$) of the six quark SU(3)-flavor
[SU(3)$_f$] multiplets \cite{su3}.  This hypercharge ($Y$) = 0 dibaryon
was predicted by Jaffe \cite{jaffe} to be stable against strong decay, but
not to weak decay; hence the experiments generally searched
for characteristic weak decay topologies.  So far, no conclusive
evidence for a bound H$_0$ has been presented
\cite{sakai,kling,HexpKK,shah,belz}.
However, the H$_0$ mass may be larger than 
$M_{\Lambda \Lambda} \equiv M_\Lambda + M_\Lambda$, the $\Lambda \Lambda$
strong decay threshold \cite{bash,negele,Hmass}. In this case
the H$_0$ could be a strong interaction resonance.
If 
the resonance lies only a few MeV above $M_{\Lambda \Lambda}$,
then it may be possible to see experimental evidence 
for its existence\footnote{
In the meson sector nature provides several examples of narrow states that lie
a few MeV above their strong decay thresholds. 
The $\phi (1020)$, $D^*(2010)^{\pm}$, $D^*(2007)^0$ and $D_{\rm S1}(2536)^+$
mesons are 32.1, 5.8, 7.1 and 35.0~MeV, respectively, above their strong
decay thresholds and have total widths of 4.4, $<$0.13, $<$2.1 and $<$2.3~MeV,
respectively.  The latter three quantities represent upper limits at the 90\%
confidence level.} \cite{bash}.  
In addition, it is important to point out
that several other doubly-strange dibaryon states occur in the SU(3)$_f$
multiplets which do not directly strong decay to $\Lambda \Lambda$ and
which may also appear as resonances in the
$\Lambda \Lambda$ and $N\Xi$ systems.

Experimental searches for a stable H$_0$ have attempted to create
and observe
the H$_0$ through: (1) double-strangeness-exchange reactions, such as
$A(K^-,K^0)X$, $A(K^-,K^+)X$; (2)  $\Xi^- A$ capture \cite{HexpKK} on
nuclear targets with $A \geq 2$; and (3)  
proton + nucleus \cite{shah,belz} or
nucleus + nucleus collisions where the H$_0$ production mechanism could
proceed via multi-hyperon production followed by coalescence
\cite{cousins,dover,sorge,kahana}. 
Reactions such as $d(K^-,K^0)$H$_0$, for example, 
produce a relatively clean final state but suffer from poor momentum
matching and subsequently, a reduced production rate.  Heavy ion collisions,
although producing a complex final state, generate a large number of
co-moving hyperons and non-strange baryons from which dibaryons could either
coalesce or scatter through resonances.
The invariant mass
spectrum for
$\Lambda \Lambda$ pairs produced in 158 A GeV/c Pb + Pb collisions
[data from WA97 \cite{wa97}]
does not reveal any resonance like structures. However, the data are too
coarsely binned and lack sufficient statistics to be relevant for the type
of experimental search proposed here.
On the other hand, a three standard deviation enhancement above background
was observed in the $\Lambda \Lambda$ invariant mass spectrum from
threshold to 30~MeV above threshold, in the $\Lambda \Lambda$ pair production
data from the $^{12}$C$(K^-,K^+)$ reaction \cite{ahn}.

The Relativistic Heavy Ion Collider (RHIC) facility
at the
Brookhaven National Laboratory, together with four new detectors,
is beginning a program of research whose primary goal is to
produce a quark-gluon plasma and study its properties.
It is noteworthy that 
a color-deconfined QGP could enhance significantly
the production of H$_0$ dibaryons compared to the production
expected from a hot hadronic gas \cite{dover}.  
The reason is  that the
largest component of the six quark dibaryon wave function consists of a
color-singlet superposition of color-octet-octet components 
\cite{bash,dover,bicker}:
\begin{equation}
\mid\!H\rangle = \sqrt{\frac{1}{5}} \mid\!BB\rangle +
                 \sqrt{\frac{4}{5}} \mid\!8_c \otimes 8_c \rangle ~.
\label{Eq1}
\end{equation}
Here, 
$\mid\!8_c \otimes 8_c \rangle$ represents the color-octet-octet components,
and $\mid\!BB\rangle$ denotes a summation of states composed
of two physical baryons.  Enhanced dibaryon production can occur by way
of fusion of color-octet clusters within the deconfined plasma \cite{dover}.
If evidence for dibaryon resonances is
observed in the data from RHIC collisions, then
analyses of such data would provide an independent basis for studying the QGP. 

In this work 
we explore the feasibility of observing evidence for the existence
of several doubly-strange dibaryon {\em resonances}
through study of the $\Lambda \Lambda$ 
and $N\Xi$ invariant mass spectra 
constructed from relativistic heavy ion collision data to be taken
at RHIC.\footnote{Formation of hyper-nuclear baryon-baryon bound states
or slightly unbound states is also possible
({\it e.g.} strange baryon
analogs of the deuteron and the nucleon + nucleon $^1$S$_0$ unbound state).
Such states, if they exist, are expected to produce broad enhancements over
many tens of MeV in the $\Lambda \Lambda$ or $N\Xi$ spectra, and not the
narrow resonance structures predicted for six quark states which are the
focus of the present work.}
We focus on using the STAR \cite{star} detector [with the Silicon Vertex 
Tracker (SVT) and Silicon Strip Detector (SSD) \cite{ssd}
upgrades] since it is the most suited of the RHIC detectors for such a
program.
A preliminary study \cite{ssd,coffin}
suggested that STAR could be used to search
for H$_0 \rightarrow \Lambda \Lambda$ with resonance masses a few hundred
MeV above $M_{\Lambda \Lambda}$.
The focus of the
present work is on the $N\Xi$ decay channel with resonance masses a few MeV
above the strong decay threshold.

In Sec.~\ref{section:states}
the doubly-strange dibaryons  and their possible decay
schemes are discussed.
Using the
P-matrix formalism \cite{bash} the resonance widths are estimated
in Sec.~\ref{section:widths}.
In Sec.~\ref{section:FSDR} 
the $J^{\pi} = 0^+$, I(isospin)~=~1, I$_3$~=~0 dibaryon member of the
SU(3)$_f$ {\bf 27}-plet \cite{su3} [referred to as H$^{27}(J^{\pi}$,I,I$_3$)
with $J^{\pi} = 0^+$, I~=~1 and I$_3$~=~0],
which cannot directly
strong decay to $\Lambda \Lambda$, is used to explore the
statistical aspects and sensitivity
of analysis of STAR data to the existence of such resonances,
if they exist. 
A new, fast simulation detector response
code (FSDR) is used for this study.
Finally, a summary and conclusions are presented in
Sec.~\ref{section:summary}.

\section{Dibaryon States}
\label{section:states}

The direct product space of the $J^{\pi} = \frac{1}{2} ^+$ baryon octet
with itself can be written in terms of irreducible representations
of SU(3)$_f$ \cite{su3}:
\begin{equation}
{\bf 8} \otimes {\bf 8} = {\bf 1} \oplus {\bf 8} \oplus {\bf 8} \oplus
                          {\bf 10} \oplus \overline{{\bf 10}} \oplus {\bf 27}.
\label{Eq2}
\end{equation}
The  hypercharge ($Y$) ranges from +2 (NN, 
nucleon-nucleon) to $-$2 ($\Xi \Xi$) for these dibaryon states.  
The $Y=2$ members include
the deuteron of the $\overline{{\bf 10}}$-multiplet.  The $Y=1$ states 
contain N$\Lambda$ and N$\Sigma$ components.  Sharp enhancements in
the N$\Lambda$ spectra at 2129~MeV \cite{pigot} [the N$\Sigma$ threshold
\cite{piekarz,doverfesh}] and 2139~MeV \cite{piekarz}
have been observed in single
strangeness-exchange reactions such as d$(K,\pi )$X and d$(\pi,K)$X.  The
$Y$=~$-$1 and $-$2 states contain the experimentally difficult $\Xi \Lambda$,
$\Xi \Sigma$ and $\Xi \Xi$ components.  Finally, the $Y=0$ dibaryons
contain $\Lambda \Lambda$, $N\Xi$, 
$\Lambda \Sigma$, and $\Sigma \Sigma$
components.  Since $N$s, $\Lambda$s, $\Sigma$s, and 
$\Xi$s are produced in relativistic heavy ion $A + A$ central collisions,
the RHIC experiments will provide an excellent opportunity
to  search for some of these $Y = 0$ dibaryons.

Table~\ref{TableI} lists the 
$J^{\pi} = 0^+,1^+$
$Y=0$ dibaryon constituents of the various
SU(3)$_f$ multiplets in Eq.~(\ref{Eq2}) from Refs.~\cite{jaffe,su3}
along with the dominant decay modes
for a wide range of assumed dibaryon masses. 
The mass limits
where strong and/or electromagnetic (EM) decay channels
open are noted  for each state.  In general the best mass ranges to
explore for resonances
are those lying just above the strong and/or EM decay thresholds in
Table~\ref{TableI}.
Jaffe \cite{jaffe} predicted a significant
increase in the dibaryon masses with increasing dimensionality of the
SU(3)$_f$ representation.  However, more 
recent calculations based on QCD sum rules
\cite{kodama,sumrule} indicate that the $Y=0$ dibaryons in different
multiplets should be similar in mass; the principal mass dependence of the
{\bf 8}$\otimes${\bf 8} dibaryons is due to the explicit SU(3)$_f$
symmetry breaking caused by the strange quark mass.\footnote{It is also
worth noting that the $Y=2$
deuteron member of the $\overline{{\bf 10}}$-flavor multiplet and the
nucleon + nucleon $^1$S$_0$ unbound state member of the {\bf 27}-flavor
multiplet differ in mass by only a few MeV.  Calculations \cite{doverfesh}
based on a SU(3) invariant pseudoscalar and vector meson exchange model for the
baryon-baryon system predict the 0$^+$ {\bf 27}-plet dibaryon
mass to be less than the
mass of the singlet H$_0$.}  
All of the $Y=0$ dibaryon states in
Table~\ref{TableI} will be considered in this work, since at present
there is no reason to exclude them from experimental searches.

For masses between $M_{\Lambda \Lambda}$ and
$M_{N\Xi}$  the singlet H$_0$ and H$^{27}(0^+,0,0)$ 
can strong decay to $\Lambda \Lambda$.  
Recently, using a P-matrix formalism, Bashinsky and Jaffe \cite{bash}
considered 
a hypothetical 
singlet H$_0$ strong-decay resonance of mass
several
MeV above the $\Lambda \Lambda$ (strong decay) threshold.
For a variety of  assumptions, they found 
that, owing to kinematic effects near threshold,
the cross sections for  S-wave $\Lambda \Lambda$
elastic scattering should show structures whose scales are of the
order of several MeV. 
It may be that
the $\Lambda \Lambda$
invariant mass spectrum generated from relativistic heavy ion
collision data will show similar structure that can be taken
as evidence of the  existence of such a resonance. 
The analysis in \cite{bash}      
also applies for the    
H$^{27}(0^+,0,0)$ case.  

The {\bf 8}-, {\bf 10}- and
$\overline{{\bf 10}}$-plet
$J^{\pi} = 1^+$ dibaryons
and the H$^{27}(0^+,1,$I$_3$) and H$^{27}(0^+,2,$I$_3$) 
dibaryons cannot strong decay
to $\Lambda \Lambda$ since antisymmetrization requires the final 
$\Lambda \Lambda$ system to be spin-singlet-even or spin-triplet-odd and
I($\Lambda \Lambda$)=0.  The H$^{27}(0^+,1,0)$ and H$^{27}(0^+,2,0)$
dibaryons can, however, strong decay to $\Lambda \Lambda$
by way of the small I=0 isospin
admixture which contributes to the physical states due to Coulomb induced
isospin mixing, but this transition rate should be much less than that for
H$_0 \rightarrow \Lambda \Lambda$.
The four $J^{\pi} = 1^+$, I$_3$=0 states can EM decay to 
$\Lambda \Lambda \gamma$ via  E1 transitions leaving the $\Lambda \Lambda$
in relative $^3$P$_J$ states.  The H$^{27}(0^+,1,0)$ and H$^{27}(0^+,2,0)$
states can also EM decay to $\Lambda \Lambda \gamma$, for
example via E1 transition to the $^3$P$_1$ $\Lambda \Lambda$ final state.  All
remaining states ({\it i.e.,} those with I$_3 \neq 0$) in this mass range
(from $M_{\Lambda \Lambda}$ to $M_{N\Xi}$)
can  decay only weakly.
It is likely that the 
$J^{\pi} = 1^+$ states are of larger mass than the $0^+$ 
states and thus even more difficult
to deal with experimentally.

In the mass range from $M_{N\Xi}$ to $M_{\Sigma \Sigma}$ all states with
I~$<$~2 decay strongly to $\Lambda \Lambda$, $N\Xi$ or $\Lambda \Sigma$ as
indicated in Table~\ref{TableI}.  The H$^{27}(0^+,2,{\pm}1)$
[H$^{27}(0^+,2,0)$] states
can strong decay to $N\Xi$ and 
$\Lambda \Sigma$ [$N\Xi$, $\Lambda \Sigma$ and $\Lambda \Lambda$]
via isospin admixtures or they can decay electromagnetically.
The H$^{27}(0^+,2,{\pm}2)$
remain blocked to both strong and EM decays up to $M_{\Sigma \Sigma}$ due to
charge conservation.
The H$^{27}(0^+,1,I_3)$ members of 
the {\bf 27}-multiplet 
cannot strong decay to $\Lambda \Lambda$ because of isospin 
conservation; however, they can strong decay to $N\Xi$ 
and $\Lambda \Sigma$ if the
mass is greater than $M_{N \Xi}$ and $M_{\Lambda \Sigma}$, respectively.  
Above $M_{\Sigma \Sigma}$ all states in 
Table~\ref{TableI} can decay strongly.

Table~\ref{TableII} lists the optimum decay channels and mass ranges 
for possible $Y=0$ dibaryon resonance searches.  The last column
indicates whether the  channels include those 
 appropriate for STAR
\cite{star}.  Of the listed decay products,
STAR can identify only protons (p), $\Lambda$s and $\Xi^-$s.
Eight of the $Y=0$ dibaryon states in 
Tables~\ref{TableI} and \ref{TableII} (corresponding to each of 
the I$_3$=0 states)
are seen to be  appropriate for 
a program of research with STAR.

Evidence for the
H$_0$ and H$^{27}(0^+,0,0)$ states might be seen
in the
$\Lambda \Lambda$ invariant mass spectrum within several MeV of threshold.
The four $J^{\pi} = 1^+$, I$_3$=0 states, if in the mass range from
$M_{\Lambda \Lambda}$ to $M_{n \Xi^0}$, will decay electromagnetically to
$\Lambda \Lambda \gamma$.
If the photon is not detected,
the $\Lambda \Lambda$ invariant mass spectrum would
appear as a broad continuum. 
However, an enhancement still might be observed
if the resonance masses are only a few MeV above
$M_{\Lambda \Lambda}$.  These four resonances might also be observed in the
$p \Xi^-$ spectrum if either lies just above the $M_{p \Xi^-}$~=~2259.6~MeV
strong breakup threshold.\footnote{Due to Coulomb interactions the $n \Xi^0$
and $p \Xi^-$ mass thresholds are split by 5.1~MeV.  Strong decay to $N\Xi$
therefore opens at $M_{n \Xi^0}$=2254.5~MeV, whereas the observable decay
channel for STAR, dibaryon$\rightarrow p \Xi^-$,
does not open until 2259.6~MeV.}

If the resonance mass exceeds $M_{\Lambda \Lambda}$,
the H$^{27}(0^+,1,0)$ and H$^{27}(0^+,2,0)$ states can decay
either electromagnetically to $\Lambda \Lambda \gamma$
or strongly via the small I=0 admixture.
The competition between these  branches determines
the experimental signature.  If the strong interaction,
isospin admixture decay
dominates the EM decay, then the I=1(2) state might be observed
if its mass is between $M_{\Lambda \Lambda}$ and $M_{n \Xi^0}$
($M_{\Sigma^0 \Sigma^0}$).  If the EM decay dominates, then both I=1 and 2
states might only be observed as enhancements in the continuum
just above $M_{\Lambda \Lambda}$.  Otherwise the H$^{27}(0^+,1,0)$
state might be seen in the $p \Xi^-$ invariant mass spectrum just above
the $M_{p \Xi^-}$ threshold.  
In this case
H$^{27}(0^+,1,0)$ $\rightarrow$ $p \Xi^-$ might be observable with
STAR.  
The H$^{27}(0^+,2,0)$
resonance could strong decay to $\Sigma^0 \Sigma^0$ and $\Sigma^+ \Sigma^-$
final states, but the $\Sigma$s, unfortunately, cannot be reconstructed
by  STAR.

In this initial simulation study we prefer to consider the dibaryon
resonance with smallest angular momentum which may strong decay
to $p \Xi^-$ but not to $\Lambda \Lambda$, namely H$^{27}(0^+,1,0)$.
The reason for this choice is the expectation that the combinatoric
background contribution to the $p \Xi^-$ 
invariant mass spectrum will be less than that for 
$\Lambda \Lambda$ for the case of RHIC data for Au + Au central
collisions (see  Sec.~\ref{section:FSDR}.) Simulation studies for the
remaining dibaryon resonances in Table~\ref{TableII} will be done
in the future.

It is worth noting that threshold-cusp
effects \cite{piekarz,doverfesh,newton} could produce
enhancements in the $\Lambda \Lambda$ spectra at the $N\Xi$, 
$\Sigma \Sigma$, etc., thresholds, independent of any possible resonances.
This is also true for the $N\Xi$ spectra (I=1 component) at the 
$\Lambda \Sigma$ and $\Sigma \Sigma$ thresholds.
Such enhancements are due to 
coupled-channels scattering effects near inelastic thresholds and must
be treated carefully
when searching for experimental evidence of dibaryon resonances.

Resonances occuring between the $n \Xi^0$ and $p \Xi^-$ thresholds will
decay primarily  to $n \Xi^0$, although decay to  $p \Xi^-$ will occur
if the resonance width is such that the distribution extends above 
$M_{p \Xi^-}$.  For resonance masses near or a few MeV above $M_{p \Xi^-}$,
the branching ratio between the $n \Xi^0$ and $p \Xi^-$  channels
depends upon the dynamics of the $n \Xi^0-p \Xi^- -$dibaryon coupled
channels system, as well as on the kinematic density of final states factor.
For resonance energies much above $M_{p \Xi^-}$ the branching ratio is 
determined mainly by isospin invariance which results in a 50\% decay fraction
for both channels.  Thus the
study in Sec.~\ref{section:FSDR}
will be in terms of the dibaryon resonance width and
production rate for the observable dibaryon$\rightarrow p \Xi^-$ decay
channel only.  Inclusion of the model-dependences associated with the
$n \Xi^0-p \Xi^- -$dibaryon system lies well beyond the scope and purpose
of this work.

Finally, we note that the
decay products of resonance states formed
deep within the interior
of the collision region will undergo strong, final state interactions (FSI)
so  that observation of these resonances
becomes more difficult. It is likely that only resonances occuring
 in the periphery of the QGP or in the
mixed QGP-hadronic phase prior to freeze-out will lead to experimental
signatures in invariant mass spectra.  Detailed calculations of FSI
effects on this and other potential QGP signals remain to be done.

\section{Dibaryon Resonance Widths}
\label{section:widths}

Bashinsky and Jaffe \cite{bash} used the P-matrix scattering formalism to
compute the $\Lambda \Lambda$ elastic scattering S-wave total cross section
in the presence of a flavor-singlet, H$_0$ dibaryon resonance located a few
MeV above the $\Lambda \Lambda$ threshold.  Their model is used in this
section for the similar case of a possible H$^{27}(0^+,1,0)$ (for brevity
referred to as H$^{27}$ in the remainder of this paper)
dibaryon resonance lying just above the $N\Xi$
threshold.

The total cross section for $N\Xi$ elastic scattering is given by
\begin{equation}
\sigma_{\rm TOT} = \frac{4\pi}{k_1} {\rm Im}(f_{11})~~~~,
\label{Eq3}
\end{equation}
where $f_{11}$ is approximated by the forward, S-wave elastic scattering
amplitude, which in the P-matrix formalism is written as \cite{bash}
\begin{equation}
f_{11}=\frac{1}{\frac{1}{K_1^{(red)}} - ik_1}.
\label{Eq4}
\end{equation}
The reduced K-matrix from Ref.~\cite{bash} as a function of total energy
($\epsilon$) in the $N\Xi$ channel is
\begin{equation}
K_1^{(red)}(\epsilon) \simeq -\overline{a}_1^\prime - 
\frac{\rho_1^\prime / 2m}{\epsilon - \epsilon_{th1} - \epsilon_1^\prime}~~~,
\label{Eq5}
\end{equation}
where,
\begin{equation}
\overline{a}_1^\prime = \frac{\overline{a}_1}{1 - \rho_1b_0},
\rho_1^\prime = \rho_1 \frac{1 - \rho_1b_0 +
2m\epsilon_1\overline{a}_1b_0}{(1 - \rho_1b_0)^2},
\epsilon_1^\prime = \frac{\epsilon_1}{1 - \rho_1b_0}~~~.
\label{Eq6}
\end{equation}
In Eqs.~(\ref{Eq3})-(\ref{Eq6}), $k_1$ is the relative center-of-momentum system
wave number, $m$ is the reduced mass of the two-baryon ($N\Xi$) system,
$\epsilon_{th1}$ is the $N\Xi$ threshold energy,
$\epsilon_1 = \epsilon_r - \epsilon_{th1}$, and 
$\epsilon_r$ is the 
assumed resonance mass of the H$^{27}$ dibaryon.
Parameter $b_0$ is a resonance size parameter with typical
value (150~MeV)$^{-1}$, $\rho_1$ is the resonance width parameter which is
a function of the resonance state - decay channel coupling, and
$\overline{a}_1$ is the $N\Xi$ scattering length which was assumed to be
(200~MeV)$^{-1}$.

Elastic total cross section results obtained
using Eqs.~(\ref{Eq3})-(\ref{Eq6}) are
shown in Fig.~\ref{figure1} by the solid lines for assumed values of
$\epsilon_1$~=~2, 3 and 5~MeV
and for the typical resonance width parameter
$\rho_1$=50~MeV.  The unitarity limit, $4\pi /k_1^2$, is indicated by the dashed
lines.  These results indicate that the resonance distribution displays a
characteristic width roughly
proportional to the difference between the resonance
and threshold energies.

Resonances decaying to $p \Xi^-$ may also decay to $n \Xi^0$ which is 
5.1~MeV lower in energy.  These results suggest that resonances near
$M_{p \Xi^-}$ would be about 5~MeV wide.  It is reasonable to assume, however,
that the $p \Xi^-$ spectra for such cases would display a narrower range of
enhancement because of the kinematic limit which cuts off the lower energy
portion of the $p \Xi^-$ invariant mass distribution.

\section{Search for Dibaryon Resonances using STAR}
\label{section:FSDR}

An environment conducive to the
formation of dibaryon resonances may be the hot, dense nuclear matter
that will be created  through central
Au + Au collisions at RHIC.  The experimental objective, then, is to 
infer evidence for the existence of such resonances
through examination of 
$\Lambda \Lambda$ or $p \Xi^-$ invariant mass spectra.  Since the 
resonance cross sections are expected to be small,
a large acceptance detector
with the ability to do precision tracking to within a few cm of 
the production vertex is needed.  In addition, the detector must
handle high data
rates and  have event processing capabilities for 
very large data volumes.  The
STAR detector \cite{star} at RHIC with its central vertex tracking system
(SVT-SSD) is well-suited for such experiments.
For the simulations to be discussed here, the parameters in 
Table~\ref{TableIII} were used to approximate the
kinematic
acceptance and reconstruction performance of STAR (including the
SVT-SSD).
The momentum resolution and track
reconstruction efficiencies were taken from previous
simulation studies \cite{ssd,sim}.

Coalescence model calculations \cite{dover,kahana}
for Au+Au central collisions at
AGS energies ($\sim$11 A$\cdot$GeV fixed target)
predict a wide range of production rates for the flavor-singlet
H$_0$ from of order $10^{-4}$ to 0.5 per event,
depending on the assumed temperature and
strangeness content of the hadron gas and on the assumed $\Lambda \Lambda$
interaction model.  Color-deconfinement in the QGP
could enhance the production rate via the color-octet-octet
component of the dibaryon wave function [Eq.~(\ref{Eq1})]. 
At RHIC energies increased strangeness production and possible QGP formation
could further enhance the dibaryon production rate above this range.
Source expansion and collective flow in RHIC collisions could also
affect coalescence model predictions \cite{sorge}.  Comprehensive
calculations including all these effects have not been done.  Nevertheless,
the results of 
such calculations suggest the necessary level of sensitivity for the
experimental search to be meaningful.

In the following, an analytical model is presented which relates the
statistical significance of the dibaryon resonance signal to the number
of Au+Au central RHIC collision events analyzed and the assumed resonance
production rate and width.  This is
followed by a description of a fast numerical simulation model and
presentation of statistical and sensitivity
results for experimental detection of a
hypothetical, narrow dibaryon resonance in the $p \Xi^-$ channel just above
threshold.

\subsection{Analytical Model}
\label{subsection:analyticalmodel}

The integrated signal for the H$^{27} \rightarrow p \Xi^-$ invariant mass
distribution can be approximated by
\begin{equation}
S \cong \sum_{i=1}^{N_{ev}} n_H^i e_p e_{\Xi} f_{H-cut}
= n_H e_p e_{\Xi} f_{H-cut} N_{ev} \equiv \bar{S} N_{ev} ~,
\label{Eq7}
\end{equation}
where $n_H^i$ is the assumed number of dibaryon resonances in the $i^{th}$
Au+Au central collision event that decay to $p \Xi^-$ with all daughter
products going fully into the kinematic acceptance range specified in 
Table~\ref{TableIII}, and $n_H$ is the average number of
H$^{27} \rightarrow p \Xi^-$ decays per event for the data sample consisting
of $N_{ev}$ events.  Also in Eq.~(\ref{Eq7}) $e_p$ and $e_{\Xi}$ are the
proton and $\Xi^-$ reconstruction efficiencies, respectively (number 
correctly reconstructed and accepted divided by number in the kinematic
acceptance or detector fiducial volume), $f_{H-cut}$ is the fraction of
reconstructed H$^{27} \rightarrow p \Xi^-$ resonance decays that survive
any remaining analysis cuts ({\it e.g.} cuts to remove
fluctuations from the background subtracted spectra and the upper/lower
$p \Xi^-$ invariant mass limits), and $\bar{S}$ is the average signal 
per event \cite{fHcut}. 
The total background underneath the signal peak is
approximated by
\begin{equation}
B \cong \sum_{i=1}^{N_{ev}} (N_p^i N_{\Xi}^i - n_H^i e_p e_{\Xi}) F(\rho_1)
\equiv \bar{B} N_{ev} ~,
\label{Eq8}
\end{equation}
where $N_p^i$ and $N_{\Xi}^i$ represent the number of reconstructed and
accepted proton and $\Xi^-$ candidates for the $i^{th}$ event, $F(\rho_1)$
represents the relative fraction of random $p \Xi^-$ pairs whose invariant
mass occurs within the domain of the H$^{27}$ mass peak, and $\bar{B}$ is
defined as the average background per event.  The fraction, $F(\rho_1)$,
depends on the proton and $\Xi^-$ momentum space distributions for the
collision events (after cuts) and on the $p \Xi^-$ invariant mass range
which in turn depends on the width of the resonance peak.

The numbers of candidate protons and $\Xi^-$ include several contributions
which are represented by the following:
\begin{eqnarray}
N_p^i & = & (n_H^i + n_{H,p}^i + n_{th,p}^i) e_p + N_{False,p} \label{Eq9} \\
N_{\Xi}^i & = & (n_H^i + n_{H,\Xi}^i + n_{th,\Xi}^i) e_{\Xi} + N_{False,\Xi}~,
\label{Eq10}
\end{eqnarray}
where $n_{H,p}^i$ ($n_{H,\Xi}^i$) is the number of protons ($\Xi^-$s) for the
$i^{th}$ event produced by H$^{27}$ resonance decays for which the proton
(all decay products of the $\Xi^-$) entered the detector acceptance but all 
decay products of the $\Xi^-$ (the proton) did not.  In Eqs.~(\ref{Eq9}) and
(\ref{Eq10}) $n_{th,p}^i$ and $n_{th,\Xi}^i$ represent the remaining
number of protons and $\Xi^-$s produced in the collision, for example by
hadronization and/or rescattering processes, which enter the detector
acceptance.  The quantities $N_{False,p}$ and $N_{False,\Xi}$ represent 
the number of incorrectly identified and accepted protons and $\Xi^-$s
per event.  The numbers of proton and $\Xi^-$ contaminants depend on various
analysis cuts, overall event multiplicity, etc.  In general, tighter cuts in
the reconstruction and analysis result in lower efficiency and reduced
contamination, while relaxed cuts have the opposite effects.  The
efficiencies and contaminants per event in this section are averages
of inclusive quantities.

The total number of counts, or yield ($Y$), for the invariant mass spectra in
the domain of the H$^{27}$ resonance peak for $N_{ev}$ Au+Au collision events
is
\begin{equation}
Y = S_T + B = S + S^{\prime} + B = (\bar{S} +\bar{S}^{\prime} + \bar{B})N_{ev},
\label{Eq11}
\end{equation}
where
\begin{equation}
S_T = \sum_{i=1}^{N_{ev}} n_H^i e_p e_{\Xi} =
n_H e_p e_{\Xi} N_{ev} = S + S^{\prime} 
\label{Eq12}
\end{equation}
which defines $S^{\prime}$.  The total error in the signal is assumed to be a
sum of statistical and systematic errors given by
\begin{equation}
\Delta S = \Delta S_{stat} + \Delta S_{syst} ~,
\label{Eq13}
\end{equation}
where the statistical error in $S$ is determined by the independent
statistical errors in the total yield ($\Delta Y$), background ($\Delta B$),
and $S^{\prime}$ ($\Delta S^{\prime}$), where
\begin{equation}
\Delta S_{stat} = \sqrt{(\Delta Y)^2 + (\Delta S^{\prime})^2
                +       (\Delta B)^2}
                = \sqrt{S + 2(B + S^{\prime})}~.
\label{Eq14}
\end{equation}
The systematic error in $S$ is assumed to be proportional to $N_{ev}$, where
$\Delta S_{syst} = \Delta \bar{S}_{syst} N_{ev}$.
 
The statistical significance of the observed resonance peak can be expressed
as the ratio of signal to signal error, which in the present model is
given by
\begin{equation}
\frac{S}{\Delta S} = \frac{\bar{S} N_{ev} } {\sqrt{\bar{S} + 2(\bar{B} +
\bar{S}^{\prime})} \sqrt{N_{ev}} + \Delta \bar{S}_{syst} N_{ev} } ~.
\label{Eq15}
\end{equation}
If $f_{H-cut}$ is constant \cite{fHcut} and
systematic errors are negligible, $S/\Delta S$ increases with
$\sqrt{N_{ev}}$ as expected, whereas for finite $\Delta \bar{S}_{syst}$,
$S/\Delta S$ is limited to $\bar{S}/\Delta \bar{S}_{syst}$ for many
events.

Solving Eq.~(\ref{Eq15}) for $N_{ev}$ yields (assuming constant $f_{H-cut}$
\cite{fHcut})
\begin{equation}
N_{ev} = \frac{(S/\Delta S)^2 [\bar{S} + 2(\bar{B} +\bar{S}^{\prime})]}
{[\bar{S} - (S/\Delta S) \Delta \bar{S}_{syst} ]^2 }
\label{Eq16}
\end{equation}
which provides an estimate of the number of collision events which must be
obtained and analyzed in order to achieve a specified value of
$S/\Delta S$.  Systematic errors limit the detection capability such that
\begin{equation}
\bar{S}  >  (S/\Delta S) \Delta \bar{S}_{syst}~,
\label{Eq17}
\end{equation}
which results in a minimum, detectable dibaryon resonance production rate
given by,
\begin{eqnarray}
n_H & > & \frac{(S/\Delta S) \Delta \bar{S}_{syst} }{e_p e_{\Xi} 
f_{H-cut}} ~,
\end{eqnarray}
assuming the required number of events from Eq.~(\ref{Eq16}) can be 
achieved.

Optimization of a particular experiment and data analysis program ({\it e.g.}
number of events and cut parameters) can be guided by the results presented
in this section provided a reasonable estimate of systematic errors can be
made.  The latter must wait until experience is gained with analysis of actual
STAR data and therefore lies well beyond the scope of the present study.
The remaining discussion and results focus on the statistical requirements
for detecting H$^{27} \rightarrow p \Xi^-$.  In the following, 
Eqs.~(\ref{Eq15}) and (\ref{Eq16}) were used to estimate the dependence
of $(S/\Delta S)$ on $N_{ev}$ and $n_H$.

\subsection{Numerical Simulations}
\label{subsection:numerical}

A fast simulation detector response code (FSDR) was developed to provide rapid
evaluation of the capabilities of relativistic
heavy ion experiments with respect to
proposed new measurements and research programs.
In simplest terms, FSDR projects a given input list of particles from
an event generator directly to the final, reconstructed particle list.
Specific detector acceptance and track reconstruction performance parameters
are supplied from realistic simulations and analyses.
The values assumed for STAR are listed in
Table~\ref{TableIII} \cite{ssd,sim,sn233}.  FSDR propagates (freely)
and decays unstable particles such as $\Lambda$ and $\Xi^-$
according to known branching ratios and
lifetimes and applies acceptance cuts, track finding efficiencies, particle
identification efficiencies, and momentum resolution
smearing to the charged primary particles and to the charged daughter particles
from the decays.

Following this, FSDR reconstruction of $\Lambda$ and $\Xi^-$ is very
similar to that used in the actual STAR reconstruction analysis.  The charged
particles selected for this analysis were assumed to be reconstructed in
both the SVT-SSD and TPC (Time Projection Chamber).
The SVT-SSD provides excellent
track position resolution (few tens of microns) \cite{ssd,sim,sn233}
in the decay
region within several cm of the primary vertex.  Reconstructed trajectories
(assumed to be helices)
were projected to the primary vertex where an impact parameter cut
(1~mm) differentiated primary particles ({\it i.e.,}
those assumed to emerge from the primary collision vertex) from secondary
particles ({\it i.e.,} those assumed to come from decay vertices).
Candidate $\Lambda$ decays were found by a distance of closest approach
(DCA) cut (2~mm) among the
projected trajectories of the secondary
protons and $\pi^-$s.  The resulting $\Lambda$ reconstruction in the $p\pi^-$
invariant mass spectrum is shown in Fig.~\ref{figure2} for 600 central 
Au + Au {\sc hijet} \cite{hijet}
collision events which included an average of 6.6
H$^{27} \rightarrow p \Xi^- \rightarrow p(\Lambda \pi^-)$
embedded decays per event (see following
discussion).  The cross-hatched region indicates the number of reconstructed
$p-\pi^-$ pairs actually produced by $\Lambda$ decays in the simulation.  The
reconstruction efficiency for primary $\Lambda$s ({\it i.e.} number
correctly reconstructed divided by number in acceptance) obtained here was about
5-6\%,\footnote{The
reconstruction efficiency for $\Lambda$s from $\Xi^-$ decays is somewhat
higher than that for primary $\Lambda$s
since the decay vertices are distributed farther from the
primary vertex where the proton and $\pi^-$ daughters are not as likely to
be removed by the 1~mm impact parameter cut as occurs for the daughter
tracks from primary $\Lambda$ decays.}
in reasonable agreement with the 8\% value in Ref.~\cite{sn233}
and the range of $4-15$\% in Ref.~\cite{ssd}.
These comparisons verify that the FSDR $\Lambda$
simulation results presented here
are consistent with those obtained from more detailed studies.
The $\pm 3 \sigma$ width of the
reconstructed $\Lambda$ peak in Fig.~\ref{figure2} is about 8~MeV which
quantitatively agrees with that found in Ref.~\cite{sn233}.  Further cuts
could reduce the background for primary $\Lambda$s as in Ref.~\cite{sn233},
but this was not done here in order to retain the secondary $\Lambda$s from
$\Xi^-$ decays.

A DCA cut (4~mm) among all pairs of selected $\Lambda$ candidates [{\it i.e.,}
those with impact parameter less than 2~cm from the primary vertex and which
have a reconstructed mass within $\pm 3 \sigma$ ($\pm$4~MeV) of the peak in
Fig.~\ref{figure2}] with all remaining secondary $\pi^-$s yielded the
$\Xi^-$ spectrum shown in Fig.~\ref{figure3} using the same events as in
Fig.~\ref{figure2}.  The
cross-hatched region indicates $\Lambda$ and $\pi^-$ reconstructed pairs that
originate from $\Xi^-$ decays.  The FSDR $\Xi^-$ reconstruction efficiency
with the preceding cuts was about 0.06 which is larger than that in
Ref.~\cite{sn233}.  
The higher
reconstruction efficiency obtained here is offset
by the smaller peak signal-to-background
ratio here ($\sim \frac{1}{4}$) compared to the larger value
($\sim$~1) in Ref.~\cite{sn233}.
The $\pm 3 \sigma$ width of the reconstructed $\Xi^-$ peak in
Fig.~\ref{figure3} is about 10~MeV, in good agreement with Ref.~\cite{sn233}.
We refer to the present set of cuts as ``relaxed'' while those in 
Ref.~\cite{sn233} are ``tight.''
A tight selection of primary protons (impact parameter
from primary vertex $\leq 100 \mu$m) and $\Xi^-$ candidates (within $\pm 6$~MeV
of the peak in Fig.~\ref{figure3}) with projected DCAs $\leq 6$~mm, provided
the set of $p \Xi^-$ pairs used in the
invariant mass plot in Fig.~\ref{figure4}.

Variable numbers of 
H$^{27}$ resonances were randomly embedded
in the {\sc hijet} \cite{hijet}
event generator output for 200~A$\cdot$GeV
Au+Au central collisions according to the following distribution,
\begin{equation}
\frac{dN}{dp^3} = A {\rm e}^{-m_T/T} {\rm e}^{-\beta m_T \, {\rm cosh}(y)} .
\label{Eq19}
\end{equation}
In Eq.~(\ref{Eq19}) $m_T = \sqrt{M(H^{27})^2 + p^2_T}$ is the transverse
mass, $T$ = 238~MeV, $\beta$ = 3.0~GeV$^{-1}$, $A$ is a normalization constant,
and at midrapidity ($y$=0) the $m_T$ distribution corresponds to an effective
temperature of 139~MeV.  For this distribution function approximately 39\%
of the total H$^{27}$s have all four decay particles fully contained within
the acceptance (Table~\ref{TableIII}).
In the following discussion, 
production rates refer to the
number of H$^{27} \rightarrow p \Xi^-$ decays per event
for which all four decay particles are within the STAR
acceptance.

In FSDR the H$^{27}$ masses were randomly distributed according to the
P-matrix resonance mass distribution (Sec.~\ref{section:widths}).  The latter
was obtained from the $N\Xi$ total elastic cross section in Eq.~(\ref{Eq3})
by eliminating the non-resonant amplitude
($\bar{a}_1$ was set to 0) and by
removing the incident flux factor in the definition of the total cross
section ({\it i.e.} by multiplying by $k_1/m$).  The resonant $p\Xi^-$
invariant mass probability distribution was therefore assumed to be
\begin{equation}
P(M_{{\rm H}^{27}}) = P_0 k_1 \sigma_{\rm TOT}(\epsilon,\bar{a}_1 = 0)/m~,
\label{Eq20}
\end{equation} 
where $M_{{\rm H}^{27}} \equiv \epsilon$ is the randomly sampled H$^{27}$
mass and $P_0$ is a normalization constant.

In this work the H$^{27}$ resonance energy, $\epsilon_1$, was assumed to be
2~MeV above the $p\Xi^-$ threshold and a range of width parameters,
$\rho_1$, was assumed where $\rho_1$ = 10, 23, 37 and 50~MeV, corresponding
to mass distributions with full width at half maximum (FWHM) values of
0.94, 2.4, 4.1 and 5.6~MeV, respectively.  Average H$^{27}$ production
rates ($n_H$) of 3.1 (8 total) and 4.7 (12 total) per central Au+Au
{\sc hijet} event were assumed.  For each of the resulting eight cases,
which correspond to different values for $\rho_1$ and $n_H$, 1940 events
were analyzed.

The preceding H$^{27}$ production rates and number of events were selected
in order to generate statistically significant resonance signals, with
modest computing requirements, such that straightforward background
subtraction and signal determination techniques would suffice.  The
analytical model in the preceding subsection, when supplemented with the
present FSDR results, can be used to obtain estimates of STAR's detection
sensitivity and data volume requirements for much smaller dibaryon resonance
production rates.

The $p\Xi^-$ invariant mass spectrum is shown in Fig.~\ref{figure4} for
1940 Au+Au central collision events for the case in which $\rho_1$~=~23~MeV
(FWHM = 2.4~MeV) and $n_H$~=~3.1.  The H$^{27}$ peak is quite apparent
for this case when the resonance occurs near threshold and the background
is rapidly decreasing.  The contributions of the large number of false
$\Xi^-$s (see Fig.~\ref{figure3}) are dispersed throughout the spectrum.
Notice that momentum resolution effects do not degrade the H$^{27}$ peak.
Finite momentum resolution \cite{sim} for the final, reconstructed
daughter protons and pions in the decay chain, 
dibaryon $\rightarrow$ $p \Xi^-$ $\rightarrow$ 
$p(\Lambda \pi^- )$ $\rightarrow$ $p(p\pi^-)\pi^-$, results in about
10~MeV/c uncertainty in the $p\Xi^-$ relative momentum.  For this case,
where FWHM is 2.4~MeV, such effects would only broaden the peak in the
reconstructed $p\Xi^-$ invariant mass spectrum to about 2.6 -- 2.9~MeV.

The uncorrelated $p\Xi^-$ background distribution near threshold is
proportional to $\sqrt{\epsilon - \epsilon_{th1}}$.  Therefore, we assumed
the threshold constrained model for the $p\Xi^-$ background distribution 
given by
\begin{equation}
B_{mod}(x) = a_1 \sqrt{x} + a_2x^2 + a_3x^3 + a_4x^4 + a_5x^5,
\label{Eq21}
\end{equation}
where $x = \epsilon - \epsilon_{th1}$ and $a_1,~a_2, \ldots a_5$ are
parameters to be determined by chi-square fits.  Excellent fits were obtained
to the nonresonant backgrounds for all cases as exemplified by the fit
shown in Fig.~\ref{figure4} by the solid line.  The cross hatched portion
of the spectrum in Fig.~\ref{figure4} indicates the reconstructed 
$p\Xi^-$ pairs from actual H$^{27}$ decays where the remaining background
is seen to be in quantitative agreement with the fit.

The background subtracted peak is shown in the inset panel in
Fig.~\ref{figure4}.  For this and all other cases the mass range of the
resonance peak was taken from threshold to 7~MeV above threshold.  In 
addition, a lower cut-off of 20 counts per 1~MeV bin (specific for the
number of events used in this analysis) was applied to remove the
residual background fluctuations which are apparent in the inset panel.
The remaining counts in the 7~MeV invariant mass range, which were above the
20 counts per bin cut-off, constitute the measured signal, $S_{FSDR}$.
For this case
the H$^{27}$ reconstruction efficiency is 0.047.

The results for each case are summarized in Table~\ref{TableIV} and in
Fig.~\ref{figure5} which give the statistical significance of the signal,
$(S/\Delta S)_{FSDR}$.  The quantity $(S/\Delta S)_{FSDR}$ was calculated using
\begin{equation}
\left( \frac{S}{\Delta S} \right)_{FSDR}  =
\frac{S_{FSDR}}{\sqrt{2Y_{FSDR}-S_{FSDR}}}~,
\label{Eq22}
\end{equation}
where $Y_{FSDR}$
is the total yield in the 7~MeV wide domain of the resonance peak.  Also
listed in Table~\ref{TableIV} are estimates of the number of Au+Au events
which are needed in order to achieve $(S/\Delta S)_{FSDR} = 3$,
assuming statistical
errors only, constant $f_{H-cut}$, and using the $N_{ev}$
dependence from Eq.~(\ref{Eq15}) to
scale the FSDR results.

The decrease in $(S/\Delta S)_{FSDR}$ with increased resonance width
and fixed production rate is due to
increased signal losses ({\it i.e.} reduced $f_{H-cut}$)
where, (1) more counts are lost in the tails of
the resonance distribution above the upper mass limit, and (2) more
counts are lost due to the 20 counts per bin cut-off.  If the resonance
mass domain was also increased, then $(S/\Delta S)_{FSDR}$
would further decrease
due to the larger background contribution to $\Delta S$.  From
Table~\ref{TableIV} and Fig.~\ref{figure5} it is seen that the required
number of events to analyze in order to achieve a certain statistically
significant signal [{\it e.g.} $(S/\Delta S)_{FSDR} = 3$] increases for broader
resonances and/or reduced production rates, as expected.

For analysis of real STAR data, it is likely that a more accurate
background subtraction method will be needed than that used here, since
optimistic production rates were assumed in this numerical study. 
A possible method is to form
random $p\Xi^-$ invariant mass histograms using mixed event pairs, normalized
to the actual number of $p\Xi^-$ pairs in the data.  This removes
dynamical correlations in the $p\Xi^-$ background spectra and is analogous
to similar methods used successfully in
pion interferometry analyses \cite{hbtbkg}.

For smaller rates of production ({\it e.g.} $n_H \approx 1$) 
Eq.~(\ref{Eq16}) simplifies to,
\begin{equation}
N_{ev} \cong \frac{2(S/\Delta S)^2 n_{th,p} e_p (n_{th,\Xi} e_{\Xi} +
N_{False,\Xi}) F(\rho_1) }{(n_H e_p e_{\Xi} f_{H-cut})^2} ~,
\label{Eq23}
\end{equation}
where contributions from $\Delta \bar{S}_{syst}$ and $\bar{S}^{\prime}$
were neglected.  The required number of events is proportional to
$n_H^{-2}$ as expected \cite{fHcut}. 
For $n_H$(total)$\approx$1 and using the results in
Table~\ref{TableIV}, it is estimated that of order $10^4 - 10^5$ Au+Au
events must be analyzed to achieve a minimum dibaryon signal with
$S/\Delta S = 3$.  At STAR, nominal run-time operations should obtain
$10^6$ central Au+Au events in about two weeks.

It is likely that the more crucial limiting factor in STAR's ability to
detect dibaryon resonances, if they exist in the mass range considered
here, will be due to systematic errors.  Although the sources and
magnitudes of these types of errors are not known at this time, a simple
estimate can be made assuming the false counts due to systematic errors
scale with overall yield.  From Eq.~(\ref{Eq16}) the minimum 
detectable signal is determined by the inequality in Eq.~(\ref{Eq17})
provided $N_{ev}$ can be arbitrarily large.  If we assume 
$\Delta\bar{S}_{syst} = f \bar{B}$, where $f$ is a constant, and use
Eq.~(\ref{Eq7}) with constant $f_{H-cut}$ \cite{fHcut},
the detectable range of dibaryon resonance production
is given by
\begin{equation}
n_H > (S/\Delta S) f \hat{n}_H (\bar{B} / \hat{\bar{S}})
\label{Eq24}
\end{equation}
where $\hat{n}_H$ and $\hat{\bar{S}}$ refer to a specific
simulation result.  The
$n_H$(total)=8 FSDR results indicate that $(\bar{B} / \bar{S}) \approx 3$.
Requiring $S/\Delta S = 3$ and assuming, for example, that the
systematic errors are 1\% of the background, the minimum, detectable
dibaryon production rate is $n_H$(total)$\approx 0.7$.  Since STAR is a 
dedicated, long-term facility, discovery of dibaryon resonances of the 
type discussed here is well within reach of the experiment, provided the
systematic errors can be kept small (to a few percent) and the dibaryon
production rate is of order 1 per event or greater.

\section{Summary and Conclusions}
\label{section:summary}

The possibility that $Y=0$ dibaryon states may occur as narrow resonances,
if located in energy just above the strong interaction breakup threshold,
offers interesting new discovery opportunities for the relativistic heavy
ion physics program at RHIC, especially for the STAR experiment.  In the 
baryon octet $\otimes$ octet direct product space for $Y=0$ dibaryons we
have listed
eight possible states (corresponding to I$_3$=0) which could, in principle,
be discovered by STAR with its SVT-SSD central vertex tracking system.  If
produced, these I$_3$=0 states decay to $\Lambda \Lambda$ and/or $p \Xi^-$
channels, depending on the resonance energy.  In addition, the formation of
these resonances may be significantly enhanced in a color-deconfined
medium, such as the QGP, by way of the dominant color-octet-octet
component of the dibaryon wave function.  A significant change in the
production rate of these states, in conjunction with other QGP signals, would
provide strong corroborative evidence for QGP formation. 
High quality statistics for the
$p \Xi^-$ invariant mass spectrum can be expected
at STAR with the analysis of data from of order
several hundred thousand
central Au+Au collision events --- several days
of STAR data acquisition.

One specific $Y=0$ dibaryon state was used to
illustrate and estimate the detection and
sensitivity range of STAR.  This state, the $J^{\pi} = 0^+$, I=1, I$_3$=0
member of the {\bf 27}-plet, was 
assumed to lie within a few MeV above the $p \Xi^-$
strong breakup threshold.  The resonance distribution was estimated using the
P-matrix formalism \cite{bash}, and the detector response and event
reconstruction/analysis effects were estimated using a fast simulation,
detector response code.  This new simulation tool, FSDR,
enables rapid evaluation of relativistic heavy ion
detector/reconstruction/analysis capabilities for proposed, new physics
programs at RHIC.

The ability of the STAR experiment to detect a possible
H$^{27} \rightarrow p \Xi^-$ dibaryon resonance decay was evaluated
in terms of the statistical significance of the dibaryon signal as a
function of the H$^{27}$ production rate and width (FWHM).  The
numerical simulation results, together with an analytical model, were
used to estimate the data volume requirements for much smaller dibaryon
resonance production rates.  Improved background subtraction methods were
suggested for the case of reduced production rates; limitations due to 
systematic errors were also estimated.  The large data volume required
(of order $10^6$ central events) for rare particle searches of the type
discussed here is compatible with the nominal run plan for STAR.
The discovery of possible dibaryon resonances with
widths of order a few MeV which lie just above their strong decay
threshold is feasible at STAR, provided systematic errors are small (few
percent) and the dibaryon resonance production rate is of order 1 or
greater per Au+Au central collision.

\vspace{0.1in}
{\bf \Large Acknowledgements}

We thank Rene Bellwied, Helen Caines, John Harris, Peter Jones
and Craig Ogilvie for their comments and careful review of the manuscript.
The support of the STAR Silicon Vertex Tracker and Strangeness Physics
Working groups is acknowledged.
This research was supported in part by The
U.~S. Department of Energy Grant DE-FG03-94ER40845
and The Robert A. Welch Foundation Grant F-604.

\pagebreak

\clearpage

\begin{table}[h]
\begin{minipage}{\textwidth}
\caption{$J^{\pi} = 0^+,1^+$~~       
 $Y = 0$ dibaryon
constituents of the {\bf 8}$\otimes${\bf 8}
baryon octet -- baryon octet direct product space for each irreducible SU(3)
flavor representation. 
The dominant decay modes
for each resonance mass range are
shown. 
NN abbreviates ``mass of nucleon + nucleon,''
and similarly for the other
baryon-baryon pairs.
Weak decays which
reduce the number of strange quarks by 1 or 2 are denoted by ``$\Delta$S=1'' or
``$\Delta$S=2,'' respectively.  ``S'' denotes strong interaction decays.
``EM'' denotes electromagnetic decays.  ``S-Iso'' indicates 
strong interaction decays via the small isospin admixtures in the
physical states.  ``S$^{\prime}$'' indicates that strong interaction decays
from the I=1 states
are possible to both $N\Xi$ and $\Lambda \Sigma$.}
\label{TableI}
\vspace{8.0pt}
\begin{center}
\begin{tabular}{|cl|ccccccc|} \hline
SU(3)$_f$ & & & \multicolumn{5}{c}{Resonance Mass Range} &  \\
\cline{3-9}
Irrep. & ($J^{\pi}$,I,I$_3$)  &  NN-N$\Lambda$ & N$\Lambda$-N$\Sigma$ &
N$\Sigma$-$\Lambda \Lambda$ & $\Lambda \Lambda$-$N\Xi$ &
$N\Xi$-$\Lambda \Sigma$ & $\Lambda \Sigma$-$\Sigma \Sigma$ &
$> \Sigma \Sigma$  \\
\hline
{\bf 1} & 0$^+$ 0 0 & $\Delta$S=2 & $\Delta$S=1 & $\Delta$S=1 & S&S&S&S  \\
{\bf 8} & 1$^+$ 0 0 & $\Delta$S=2 & $\Delta$S=1 & $\Delta$S=1 &EM&S&S&S  \\
{\bf 8} & 1$^+$ 1 1 & $\Delta$S=2 & $\Delta$S=1 & $\Delta$S=1 & $\Delta$S=1 &
       S & S & S  \\
{\bf 8} & 1$^+$ 1 0 & $\Delta$S=2 & $\Delta$S=1 & $\Delta$S=1 &EM&S&
       S & S  \\
{\bf 8} & 1$^+$ 1 $-$1 & $\Delta$S=2 & $\Delta$S=1 & $\Delta$S=1 & $\Delta$S=1 &
       S & S & S  \\
{\bf 10} & 1$^+$ 1 1 & $\Delta$S=2 & $\Delta$S=1 & $\Delta$S=1 & $\Delta$S=1 &
       S & S$^{\prime}$ & S  \\
{\bf 10} & 1$^+$ 1 0 & $\Delta$S=2 & $\Delta$S=1 & $\Delta$S=1 &EM&S&
       S$^{\prime}$ & S  \\
{\bf 10} & 1$^+$ 1 $-$1 & $\Delta$S=2 & $\Delta$S=1 & $\Delta$S=1 &
       $\Delta$S=1 &
       S & S$^{\prime}$ & S  \\
$\overline{{\bf 10}}$ & 1$^+$ 1 1 & $\Delta$S=2 & $\Delta$S=1 & $\Delta$S=1
       & $\Delta$S=1 & S & S$^{\prime}$ & S  \\
$\overline{{\bf 10}}$ & 1$^+$ 1 0 & $\Delta$S=2 & $\Delta$S=1 & $\Delta$S=1 
       &EM&S&S$^{\prime}$ & S  \\
$\overline{{\bf 10}}$ & 1$^+$ 1 $-$1 & $\Delta$S=2 & $\Delta$S=1 & $\Delta$S=1 
       & $\Delta$S=1 & S & S$^{\prime}$ & S  \\
{\bf 27} & 0$^+$ 0 0 & $\Delta$S=2 & $\Delta$S=1 & $\Delta$S=1 & S&S&S&S  \\
{\bf 27} & 0$^+$ 1 1 & $\Delta$S=2 & $\Delta$S=1 & $\Delta$S=1 & $\Delta$S=1 &
       S & S$^{\prime}$ & S  \\
{\bf 27} & 0$^+$ 1 0 & $\Delta$S=2 & $\Delta$S=1 & $\Delta$S=1 &
       $\stackrel{\rm EM,}{\rm S-Iso}$ & S & S$^{\prime}$ & S  \\
{\bf 27} & 0$^+$ 1 $-$1 & $\Delta$S=2 & $\Delta$S=1 & $\Delta$S=1 &
       $\Delta$S=1 &
       S & S$^{\prime}$ & S  \\
{\bf 27} & 0$^+$ 2 2  & $\Delta$S=2 & $\Delta$S=1 & $\Delta$S=1 & $\Delta$S=1 &
       $\Delta$S=1 & $\Delta$S=1 & S  \\
{\bf 27} & 0$^+$ 2 1  & $\Delta$S=2 & $\Delta$S=1 & $\Delta$S=1 & $\Delta$S=1 &
       $\stackrel{\rm EM,}{\rm S-Iso}$ &
       $\stackrel{\rm EM,}{\rm S-Iso}$ &  S  \\
{\bf 27} & 0$^+$ 2 0  & $\Delta$S=2 & $\Delta$S=1 & $\Delta$S=1 & 
       $\stackrel{\rm EM,}{\rm S-Iso}$ &
       $\stackrel{\rm EM,}{\rm S-Iso}$ &
       $\stackrel{\rm EM,}{\rm S-Iso}$ &  S  \\
{\bf 27} & 0$^+$ 2 $-$1 & $\Delta$S=2 & $\Delta$S=1 & $\Delta$S=1 &
       $\Delta$S=1 &
       $\stackrel{\rm EM,}{\rm S-Iso}$ &
       $\stackrel{\rm EM,}{\rm S-Iso}$ &  S  \\
{\bf 27} & 0$^+$ 2 $-$2 & $\Delta$S=2 & $\Delta$S=2 & $\Delta$S=1 &
       $\Delta$S=1 &
       $\Delta$S=1 & $\Delta$S=1 & S  \\
\hline
\end{tabular}
\end{center}
\end{minipage}
\end{table}

\clearpage

\begin{table}[h]
\begin{minipage}{\textwidth}
\caption{Optimum decay channels and resonance mass ranges for 
 $J^{\pi} = 0^+,1^+$~~  
$Y=0$ dibaryon resonance
searches.}
\label{TableII}
\vspace{8.0pt}
\begin{center}
\begin{tabular}{|cl|l|c|} \hline
SU(3)$_f$ & &  & Accessible  \\
Irrep. & ($J^{\pi}$,I,I$_3$)  & Decay Channels and
Resonance Mass Ranges (MeV) & to STAR  \\
\hline
{\bf 1} & 0$^+$ 0 0 & $\Lambda \Lambda$ $\stackrel{>}{\sim}$ 2231 & Yes  \\
{\bf 8} & 1$^+$ 0 0 & $\Lambda \Lambda$ $\stackrel{>}{\sim}$ 2231;
                      $n \Xi^0$         $\stackrel{>}{\sim}$ 2254;
                      $p \Xi^-$         $\stackrel{>}{\sim}$ 2260 & Yes  \\ 
{\bf 8} & 1$^+$ 1 1 & $p \Xi^0$         $\stackrel{>}{\sim}$ 2253 &      \\
{\bf 8} & 1$^+$ 1 0 & $\Lambda \Lambda$ $\stackrel{>}{\sim}$ 2231;
                      $n \Xi^0$         $\stackrel{>}{\sim}$ 2254;
                      $p \Xi^-$         $\stackrel{>}{\sim}$ 2260 & Yes  \\
{\bf 8} & 1$^+$ 1 $-$1 & $n \Xi^-$        $\stackrel{>}{\sim}$ 2261 &      \\
{\bf 10} & 1$^+$ 1 1 & $p \Xi^0$         $\stackrel{>}{\sim}$ 2253 &      \\
{\bf 10} & 1$^+$ 1 0 & $\Lambda \Lambda$ $\stackrel{>}{\sim}$ 2231;
                       $n \Xi^0$         $\stackrel{>}{\sim}$ 2254;
                       $p \Xi^-$         $\stackrel{>}{\sim}$ 2260 & Yes  \\
{\bf 10} & 1$^+$ 1 $-$1 & $n \Xi^-$        $\stackrel{>}{\sim}$ 2261 &      \\
$\overline{{\bf 10}}$ & 1$^+$ 1 1 & $p \Xi^0$        
                                         $\stackrel{>}{\sim}$ 2253 & \\
$\overline{{\bf 10}}$ & 1$^+$ 1 0 & $\Lambda \Lambda$ $\stackrel{>}{\sim}$ 2231;
                      $n \Xi^0$         $\stackrel{>}{\sim}$ 2254;
                      $p \Xi^-$         $\stackrel{>}{\sim}$ 2260 & Yes  \\
$\overline{{\bf 10}}$ & 1$^+$ 1 $-$1 & $n \Xi^-$       
                                        $\stackrel{>}{\sim}$ 2261 & \\
{\bf 27} & 0$^+$ 0 0 & $\Lambda \Lambda$ $\stackrel{>}{\sim}$ 2231 & Yes  \\
{\bf 27} & 0$^+$ 1 1 & $p \Xi^0$         $\stackrel{>}{\sim}$ 2253 & \\
{\bf 27} & 0$^+$ 1 0 & $\Lambda \Lambda$ from 2231 to 2254;\footnote{
     If strong decay via isospin admixture dominates EM decay and resonance
     remains narrow; otherwise $\Lambda \Lambda$ for resonance mass
     $\stackrel{>}{\sim}$ 2231~MeV only.}
                       $n \Xi^0$         $\stackrel{>}{\sim}$ 2254;
                      $p \Xi^-$         $\stackrel{>}{\sim}$ 2260 & Yes  \\
{\bf 27} & 0$^+$ 1 $-$1 & $n \Xi^-$        $\stackrel{>}{\sim}$ 2261 &      \\
{\bf 27} & 0$^+$ 2 2  & $\Sigma^+ \Sigma^+$ $\stackrel{>}{\sim}$ 2379 &   \\
{\bf 27} & 0$^+$ 2 1  & $p \Xi^0$ from 2253 to 2382;\footnote{
     If strong decay via isospin admixture dominates EM decay and resonance
     remains narrow; otherwise $p \Xi^0$ for resonance mass
     $\stackrel{>}{\sim}$ 2253~MeV only.}
                        $\Sigma^+ \Sigma^0$ $\stackrel{>}{\sim}$ 2382 &   \\
{\bf 27} & 0$^+$ 2 0  & $\Lambda \Lambda$ from 2231 to 2385;$^{a}$
                        $\Sigma \Sigma$ $\stackrel{>}{\sim}$ 2385 
                      & Yes\footnote{$\Lambda \Lambda$ decay channel only.} \\
{\bf 27} & 0$^+$ 2 $-$1 & $n \Xi^-$ from 2261 to 2390;\footnote{
     If strong decay via isospin admixture dominates EM decay and resonance
     remains narrow; otherwise $n \Xi^-$ for resonance mass $\stackrel{>}{\sim}$
     2261~MeV only.}    $\Sigma^0 \Sigma^-$ $\stackrel{>}{\sim}$ 2390 &   \\
{\bf 27} & 0$^+$ 2 $-$2 & $\Sigma^- \Sigma^-$ $\stackrel{>}{\sim}$ 2395 &   \\
\hline
\end{tabular}
\end{center}
\end{minipage}
\end{table}

\clearpage

\begin{table}[h]
\begin{minipage}{\textwidth}
\caption{Kinematic acceptance and reconstruction parameters assumed in the
analytical and numerical simulations.}
\label{TableIII}
\vspace{8.0pt}
\begin{center}
\begin{tabular}{|l|l|} \hline
Parameter & Value \\
\hline
Transverse momentum acceptance       &     $p_T > 0.1$~GeV/c  \\
Pseudorapidity acceptance            &  $\mid\! \eta \!\mid \leq 1$   \\
Azimuthal acceptance                 &  $\phi = 0 \rightarrow 2\pi$   \\
Momentum resolution, $\Delta p/p$    &  2\%                           \\
TPC track reconstruction efficiency\footnote{Time Projection Chamber; main
tracking detector for STAR.}         &  90\%                          \\
SVT track reconstruction efficiency\footnote{Silicon Vertex Tracker
and Silicon Strip Detector tracking system for STAR.}
                                     &  80\%                          \\
Particle Identification ($\pi$,protons) & 100\%\footnote{Actual value not
yet available.}   \\
\hline
\end{tabular}
\end{center}
\end{minipage}
\end{table}

\clearpage

\begin{table}[h]
\begin{minipage}{\textwidth}
\caption{FSDR results for the statistical significance
$(S/\Delta S)_{FSDR}$ of the
H$^{27} \rightarrow p \Xi^-$ dibaryon resonance decay signal for total
production rates per Au+Au central event of 8 and 12 and for four
assumed values of resonance width.}
\label{TableIV}
\vspace{8.0pt}
\begin{center}
\begin{tabular}{|cccccc|} \hline
$n_H$ & $n_H$\footnote{Number of H$^{27} \rightarrow p \Xi^-$ decays
per event in which all decay products enter the STAR detector acceptance.
This is 39\% of the total for the assumed H$^{27}$ momentum distribution model.}
 & $\rho_1$   &  FWHM   &   $(S/\Delta S)_{FSDR}$\footnote{For 1940 Au+Au
central collision events.} & $N_{ev}$ for  \\
 (Total)   &     &   (MeV)  &  (MeV)  &     & 
$(S/\Delta S)_{FSDR} = 3$\footnote{Assuming $(S/\Delta S)_{FSDR}$
is proportional to $\sqrt{N_{ev}}$ as discussed in the text.}   \\
\hline
  8   &   3.1   &   10   &   0.94   &  7.9  &  280   \\
  8   &   3.1   &   23   &   2.4    &  6.4  &  426   \\
  8   &   3.1   &   37   &   4.1    &  5.6  &  557   \\
  8   &   3.1   &   50   &   5.6    &  4.1  &  1039   \\
\hline
  12   &   4.7   &   10   &   0.94   &  10.2  &  168   \\
  12   &   4.7   &   23   &   2.4    &  9.1   &  211   \\
  12   &   4.7   &   37   &   4.1    &  8.6   &  236   \\
  12   &   4.7   &   50   &   5.6    &  7.5   &  310   \\
\hline
\end{tabular}
\end{center}
\end{minipage}
\end{table}

\clearpage

\figure
\vspace{-2.0in}
\epsffile{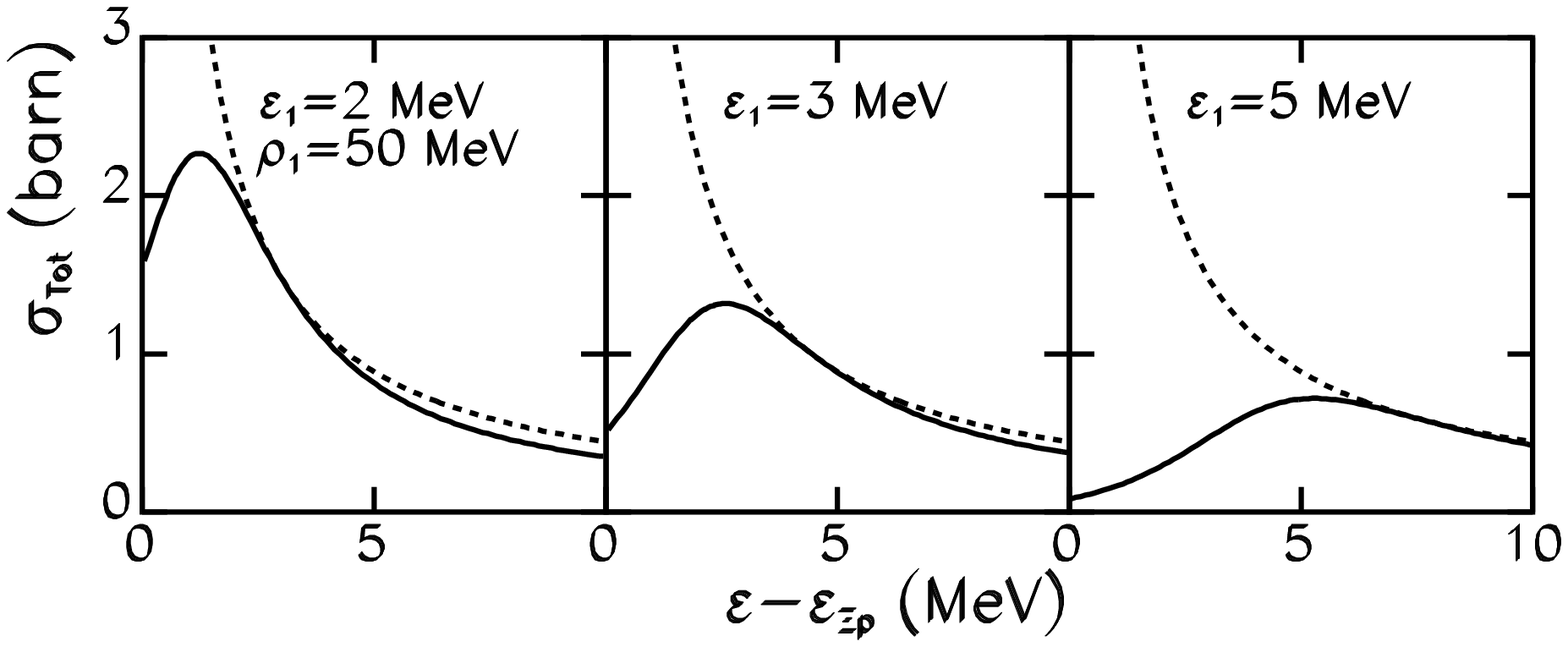}
\vspace{-3.0in}
\caption[*]{Total elastic cross sections for $N+\Xi$ scattering assuming the
H$^{27}$ dibaryon resonance is 2, 3 and 5~MeV above threshold are shown in the
left, center and right panels, respectively.  The dashed lines indicate the
unitarity limit, $4\pi /k^2_1$.  The $N\Xi$ threshold energy is denoted by
$\varepsilon_{\Xi p}$ (axis label).}
\label{figure1}
\endfigure

\clearpage

\figure
\vspace{-2.0in}
\epsffile{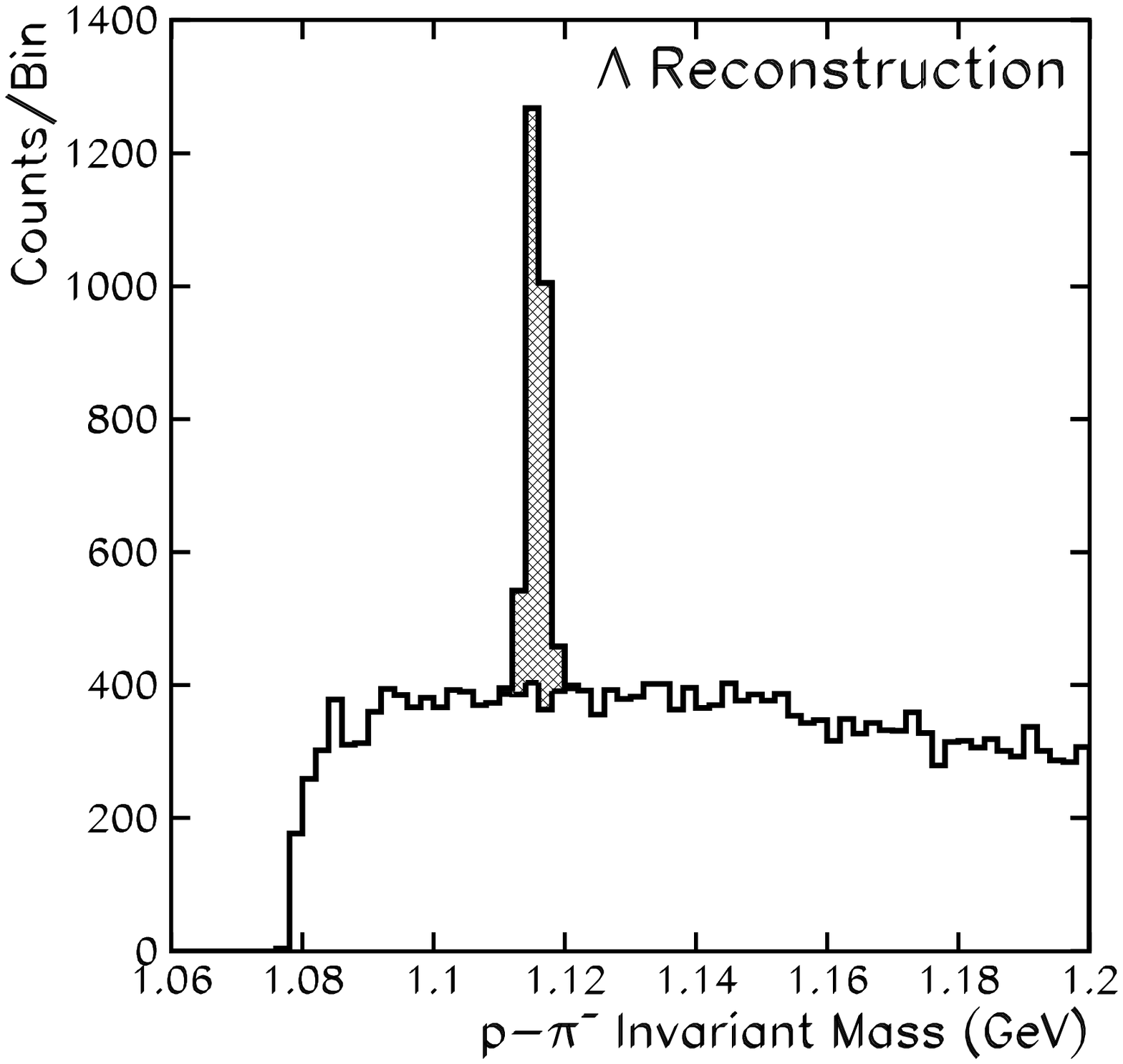}
\vspace{-1.0in}
\caption[*]{FSDR $p\pi^-$ invariant mass spectrum for $\Lambda \rightarrow
p\pi^-$ reconstruction for 600 central Au+Au {\sc hijet} events
with embedded H$^{27}$ resonances at a
collision energy of 200~A$\cdot$GeV as
described in the text.  The cross-hatched portion
indicates the number of correctly reconstructed $\Lambda \rightarrow p\pi^-$
decays.  Bin size is 2~MeV.}
\label{figure2}
\endfigure

\clearpage

\figure
\vspace{-2.0in}
\epsffile{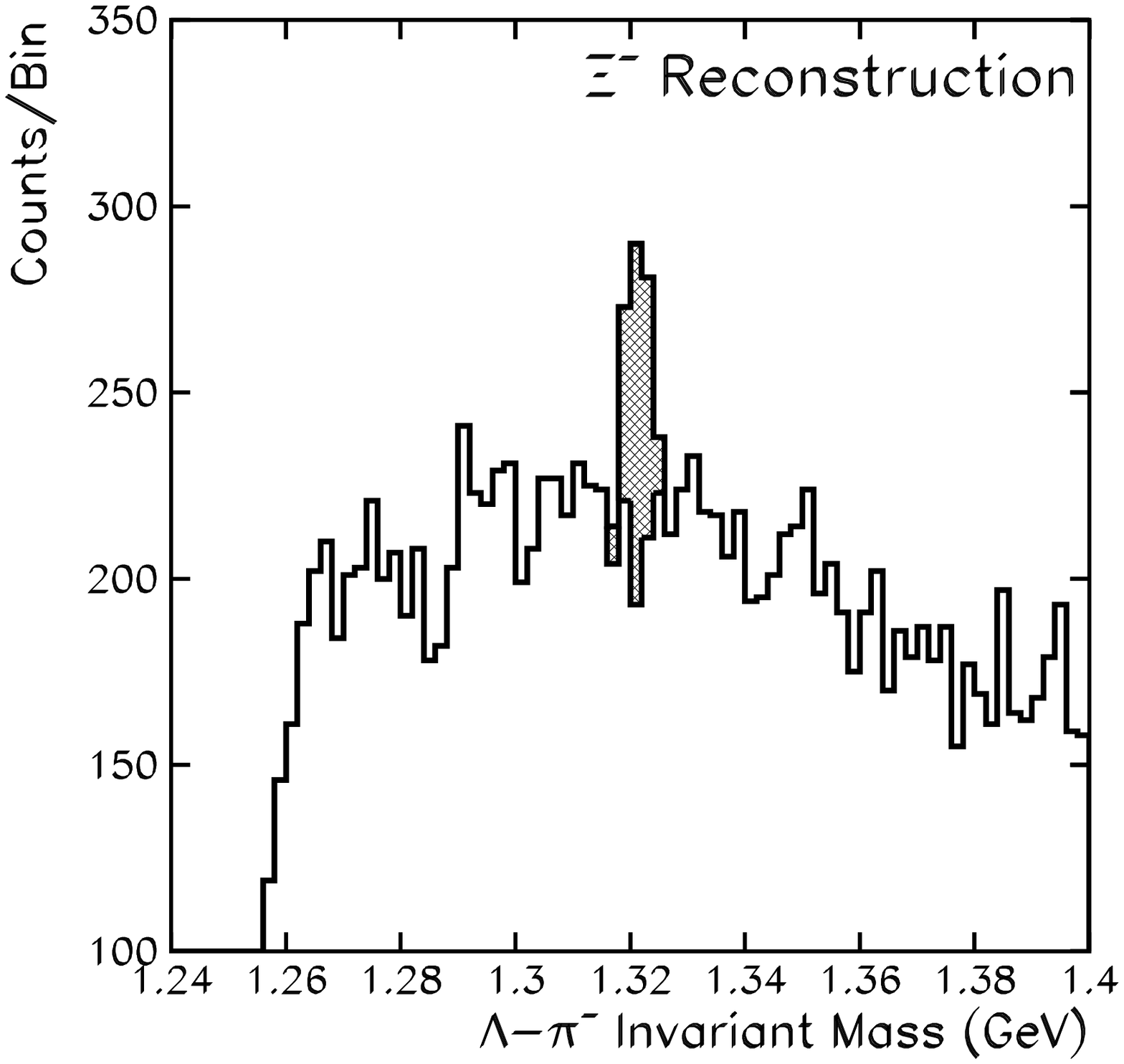}
\vspace{-1.0in}
\caption[*]{FSDR $\Lambda \pi^-$ invariant mass spectrum for
$\Xi^- \rightarrow \Lambda \pi^-$ reconstruction for 600 central Au+Au
{\sc hijet} events with embedded H$^{27}$ resonances
at a collision energy of 200~A$\cdot$GeV, using the relaxed
$\Xi^-$ decay vertex reconstruction cuts described in the text.
The cross-hatched portion indicates the number of correctly reconstructed
$\Xi^- \rightarrow \Lambda \pi^-$ decays.  Bin size is 2~MeV.  Note the
suppressed zero for the vertical scale.}
\label{figure3}
\endfigure

\clearpage

\figure
\vspace{-1.0in}
\epsffile{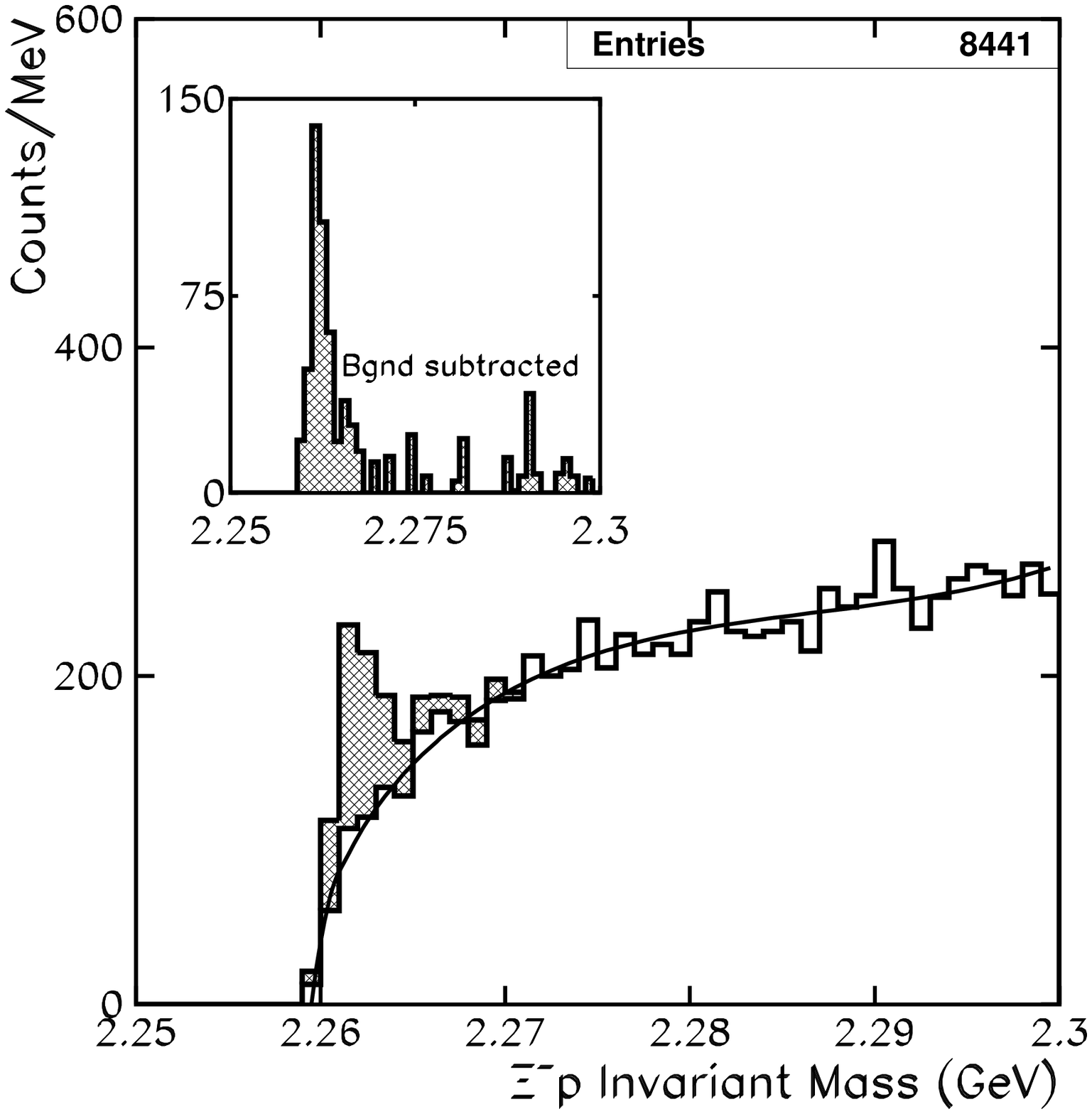}
\vspace{-1.0in}
\caption[*]{FSDR $p \Xi^-$ invariant mass spectrum for
H$^{27} \rightarrow p \Xi^-$ reconstruction for 1940 central Au+Au
{\sc hijet} events at a collision energy of 200~A$\cdot$GeV, assuming an
average of 3.1
H$^{27} \rightarrow p \Xi^-$ resonance decays per event
into the acceptance with
resonance energy 2~MeV above threshold and P-matrix width parameter
($\rho_1$) of
23~MeV (FWHM = 2.4~MeV).
The cross-hatched portion indicates the number of correctly
reconstructed H$^{27} \rightarrow p \Xi^-$ decays.  The background subtracted
peak is shown in the inset panel.  Bin size is 1~MeV.  Solid curve indicates
the five-term threshold constrained background model fit.}
\label{figure4}
\endfigure

\clearpage

\figure
\vspace{-2.0in}
\epsffile{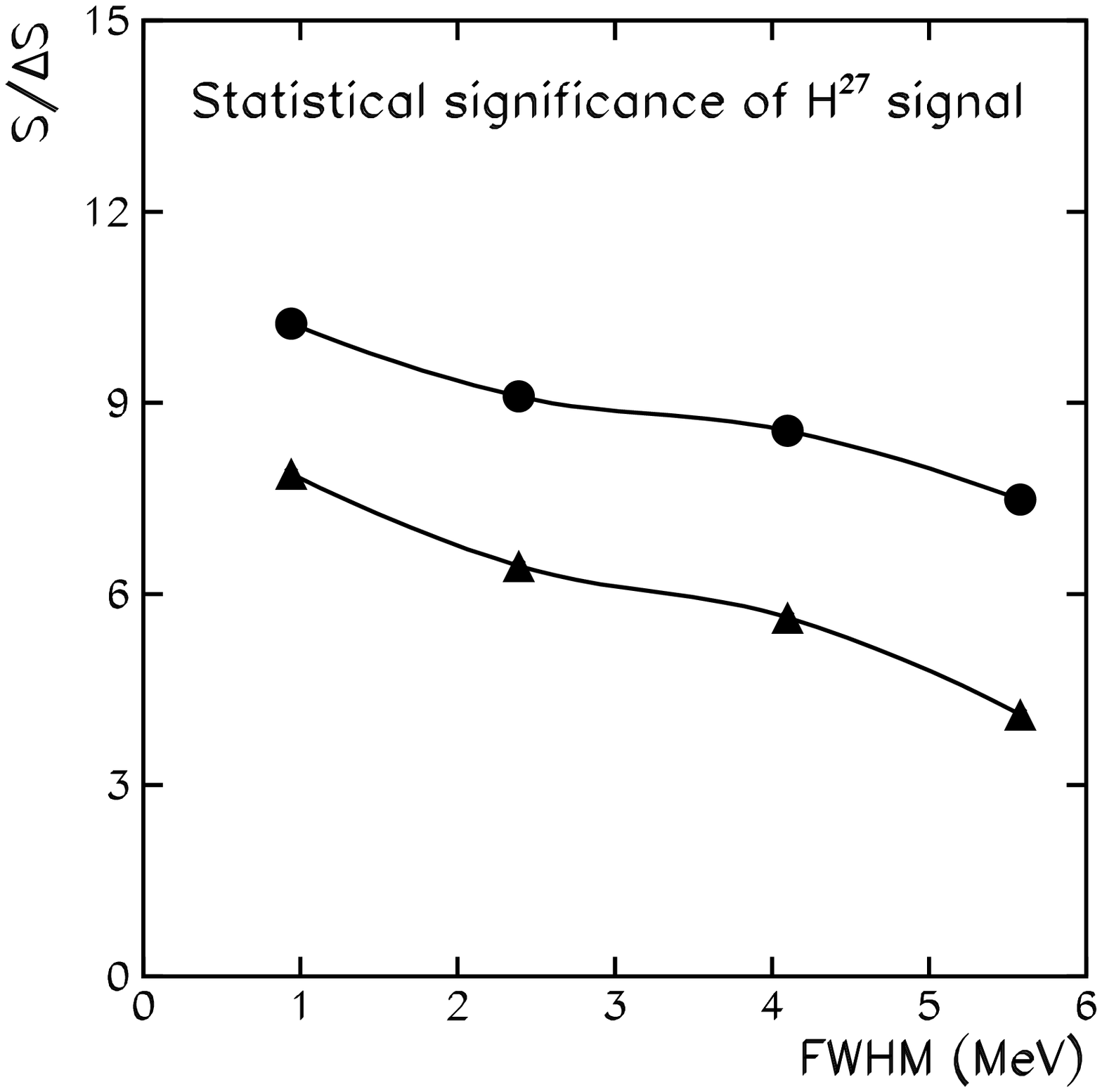}
\vspace{-1.0in}
\caption[*]{Statistical significance of the H$^{27} \rightarrow p \Xi^-$
dibaryon resonance decay signal as a function of resonance width (FWHM)
for resonance production rates of 3.1
(triangles) and 4.7 (dots) per event (all decay products enter detector
acceptance).  For each case 1940 Au+Au central collisions were analyzed.
Solid lines connect the values for fixed production rates.}
\label{figure5}
\endfigure


\begin{thebibliography}{555}
\bibitem{fHcut}
     In general $f_{H-cut}$ depends on $n_H$, $N_{ev}$, $\rho_1$, etc.
For example, the number of lost signal counts due to the background
fluctuation cut depends on the size of the statistical fluctuations in the
background subtracted spectrum which is proportional to $\sqrt{\bar{B}N_{ev}}$.
Refering to quantities in Sec.~\ref{subsection:analyticalmodel},
$f_{H-cut} \equiv S/S_T \equiv (S_T - S^{\prime})/S_T \cong
(n_H e_p e_{\Xi} N_{ev} - C\sqrt{\bar{B}N_{ev}})/(n_H e_p e_{\Xi} N_{ev})$,
where $C$ depends on the invariant mass domain size, resonance width, and
bin size.  $\bar{S}$ and $\bar{S}^{\prime}$ in general depend on $N_{ev}$.
In this work we consider two kinds of applications in which $f_{H-cut}$ can
be regarded as constant with respect to $n_H$ and $N_{ev}$: (1) strong
H$^{27}$ signals where $f_{H-cut} \approx 1$, and (2) weak H$^{27}$ signals
where $N_{ev} \propto n_H^{-2}$ is required.
\end{thebibliography}
\end{document}